\documentstyle[prl,aps,twocolumn,epsfig]{revtex}
\begin{document}
\preprint{ }
\draft
\title{\bf Spinodal Inflation}
\author{{\bf D. Cormier$^{(a)}$,
R. Holman$^{(b)}$}}
\address
{{\it (a) Centre for Theoretical Physics, University of Sussex, Falmer,
Brighton, BN1 9QH, United Kingdom} \\
{\it (b) Department of Physics, Carnegie Mellon University, Pittsburgh,
PA. 15213, U. S. A.}}
\date{December 1998}
\maketitle
\begin{abstract}
Out-of-equilibrium, non-perturbative, quantum effects 
significantly modify the standard picture of inflation in a wide
class of models including new, natural, and hybrid inflation.
We find that the quantum evolution of a single real inflaton 
field may be modeled by a classical theory of {\it two} homogeneous 
scalar fields.  We briefly discuss the important observational 
consequences that are expected to result.
\end{abstract}

\section*{}

The advent of inflation\cite{guth,inflation} is one of the most 
significant advances in
our understanding of cosmology and the early universe in the past
twenty years.  It not only provides a solution to a number of 
shortcomings of the standard model of cosmology, but also provides
the favored mechanism for the production of the primordial density 
perturbations which have been observed in the Cosmic Microwave 
Background and from which the large scale structure of our universe 
formed.

While there are any number of models of inflation,
as a paradigm it remains remarkably simple.  Even the simplest of
scalar field theory models, with appropriately chosen parameters
and initial states, are successful in producing a universe consistent
with present observation.  Furthermore, it has been thought that the 
inflaton may be treated as a purely {\it classical} field evolving in a 
{\it classical} effective potential, with quantum mechanics only 
entering in the formation of primordial density perturbations
from vacuum fluctuations\cite{mfb}, and in the uncertainty in the
inflaton initial condition.

Here we show that quantum effects can have far further ranging effects than
previously thought. Potentials with spinodal instabilities, i.e. potentials
$V(\Phi)$ for which $V''(\Phi)$ changes sign, such as those used for 
spontaneous
symmetry breaking, demonstrate a phenonmenon which we call {\it spinodal
inflation}.

In spinodal inflation, the out-of-equilibrium, non-perturbative,
quantum physics significantly modifies the evolution of the inflaton.  The
result is that the evolution of a real inflaton field may in fact be accurately
modeled not by one, but by {\it two} coupled classical fields. To do this, we
need to use the closed time path formalism of non-equilibrium quantum field
theory to study the time evolution of a real inflaton\cite{ctp}. The spinodal
instabilities give rise to non-perturbative behavior that we treat by means of
the self-consistent Hartree approximation\cite{hartree}, while the 
gravitational evolution is included by means of the semi-classical
approximation\cite{birrell}. 

Doing this we will see that the long-wavelength fluctuations assmble
themselves into an effective homogeneous {\it classical} field with only
perturbatively small quantum corrections remaining\cite{grav}. This effective 
field, together with the original zero momentum mode are the two fields that 
can be used to describe the quantum non-equilibrium evolution of the full 
system.

The resulting phenomenology is rich and complex.  Observables depend
not only on the parameters of the inflationary theory, but also on
the initial conditions.  Furthermore, observables may take on values
in the complete quantum theory that are not allowed according to  the 
simple classical analyses.  

We work in a spatially flat Friedmann-Robertson-Walker universe
with metric
\begin{equation}
ds^2 = dt^2 - a^2(t) d\vec{x}^2 \; ,
\label{metric}
\end{equation}
and we take the inflaton to be a real scalar field with Lagrangian
$$
L = \frac12 \nabla^\mu \Phi(x) \nabla_\mu \Phi(x) -
V\left[\Phi(x)\right] \; .
$$
We will be interested in the case in which the potential is even in
$\Phi$ with a negative squared mass, such that there is a local 
maximum at $\Phi=0$.

It is convenient to break up the field $\Phi$ into its expectation
value, defined within the closed time path formalism, and fluctuations
about that value:
$$
\Phi(\vec{x},t) = \phi(t) + \psi(\vec{x},t) \; , \; \;
\phi(t)  \equiv  \langle \Phi(\vec{x},t) \rangle \; .
$$
By definition $\langle \psi(\vec{x},t) \rangle = 0$, while $\phi$ 
depends only on time as a consequence of space translation 
invariance. 

By imposing the Hartree factorization, we arrive at the following
equations of motion for the inflaton\cite{thesis}:
\begin{equation}
\ddot{\phi} + 3\frac{\dot{a}}{a}\dot{\phi} + 
\sum_{n=0}^{\infty} \frac{1}{2^n n!} \langle \psi^2 \rangle^n
V^{(2n+1)}(\phi)  = 0 \; ,
\label{phieqn} 
\end{equation}
\begin{equation}
\left[\frac{d^2}{dt^2} + 3\frac{\dot{a}}{a}\frac{d}{dt} + 
\frac{k^2}{a^2} + \sum_{n=0}^{\infty} \frac{1}{2^n n!} 
\langle \psi^2 \rangle^n V^{(2n+2)}(\phi) \right] f_k = 0 \; ,
\label{fkeqn}
\end{equation}
where 
$$
V^{(n)} \equiv \frac{\delta^n V(\phi)}{\delta \phi^n} \; .
$$
The two-point fluctuation $\langle \psi^2 \rangle$ is determined 
from the mode functions $f_k$:

\begin{equation}
\langle \psi^2 \rangle = \int \frac{d^3k}{2(2\pi)^3} |f_k|^2 
\; .
\label{fluctuation}
\end{equation}
For $a(t_0)=1$, the initial conditions on the mode functions 
are
\begin{equation}
f_k(t_0) = \frac{1}{\sqrt{\omega_k}} \; , \; \;
\dot{f}_k(t_0) = \left(-\dot{a}(t_0) - 
i\omega_k \right) f_k(t_0) \; ,
\label{fkin}
\end{equation}
with
$$
\omega_k^2 \equiv k^2 + \sum_{n=0}^{\infty} \frac{1}{2^n n!} 
\langle \psi^2(t_0) \rangle^n V^{(2n+2)}(\phi(t_0))
- \frac{R(t_0)}{6} \; .
$$
$R(t_0)$ is the initial Ricci scalar.
For $k^2 < |V^{(2)}(\phi(t_0))|$ we modify $\omega_k$ either by
means of a quench or by explicit deformation so that the frequecies
are real.  The exact choice corresponds to different initial 
vacuum states and has little effect on results\cite{grav,desitter}.

The gravitational dynamics are determined by the semi-classical
Einstein equation\cite{birrell}. For a minimally coupled inflaton we have:
\begin{eqnarray}
\frac{\dot{a}^2}{a^2} &=& \frac{8\pi G_N}{3} 
\left[\frac12 \dot{\phi}^2 + \frac12 \langle \dot{\psi}^2 \rangle
+ \frac{1}{2a^2} \langle (\vec{\nabla}\psi)^2 \rangle \right. \nonumber \\
&+& \left.\sum_{n=0}^{\infty} \frac{1}{2^n n!} 
\langle \psi^2 \rangle^n V^{(2n)}(\phi)\right] \; ,
\label{frweqn}
\end{eqnarray}
where $G_N$ is Newton's gravitational constant, and 
\begin{eqnarray}
\langle \dot{\psi}^2(t) \rangle &\equiv& \int \frac{d^3k}{2(2\pi)^3}
|\dot{f}_k|^2  \; , 
\label{psidot} \\
\left\langle \left(\vec{\nabla} \psi(t) \right)^2 \right\rangle &\equiv&
\int \frac{d^3k}{2(2\pi)^3} k^2 |f_k|^2 \; .
\label{nablapsi}
\end{eqnarray}
In what follows, we assume that each of these integrals has been regulated
either because the theory is a low energy effective theory with a definite
cutoff or because the divergences have been absorbed into a renormalization of
the parameters of the theory\cite{grav,baacke}.

We now describe zero mode assembly\cite{grav}.  
We write the potential $V(\Phi)$ as
\begin{equation}
V(\Phi) = K - \frac12 \mu^2 \Phi^2 
+ \frac{\lambda}{4!}\Phi^4 + \cdots \;,
\label{pot2}
\end{equation}
where the constant $K$ is chosen such that the potential is zero
in the true vacuum.  Initially, the fluctuations $\langle \psi^2
\rangle$ are small, so for $\phi_{\rm initial} \ll \mu/\sqrt{\lambda}$,
the mode functions evolve as
$$
\ddot{f}_k + 3H_0 \dot{f}_k +\left(\frac{k^2}{a^2}-\mu^2\right)f_k=0
\; ,
$$
where $H_0^2 \simeq 8\pi G_N K/3$.  As a result, those modes whose physical
wavelength is greater than the horizon scale with $k/a < H_0$ grow
exponentially; this is the spinodal instability. After a few $e$-folds of
inflation, the integral (\ref{fluctuation}) becomes dominated by long
wavelength modes, which are the most unstable ones, so that we may replace the
quantity $\langle \psi^2(t) \rangle^{1/2}$ by an effectively {\it homogeneous}
and classical zero mode $\sigma(t)$.  Furthermore, the gravitational evolution
may be written in terms of $\sigma$ with the replacements $\langle \dot{\psi}^2
\rangle \to \dot{\sigma}^2$ and $\langle (\vec{\nabla}\psi)^2 \rangle/a^2 \to
0$.  Both of these replacements are justified once a few $e$-folds of inflation
have passed\cite{grav}.

Finally, we arrive at the following system of equations for 
$\phi$ and $\sigma$:
\begin{eqnarray}
\ddot{\phi}+3\frac{\dot{a}}{a} \dot{\phi} + \sum_{n=0}^{\infty}
\frac{1}{2^n n!} \sigma^{2n} V^{(2n+1)}(\phi) & = & 0 \; , 
\label{assembly1} \\
\ddot{\sigma} + 3\frac{\dot{a}}{a} \dot{\sigma} + \sum_{n=0}^{\infty}
\frac{1}{2^n n!} \sigma^{2n+1} V^{(2n+2)}(\phi) & = & 0 \; , 
\label{assembly2}
\end{eqnarray}
while
\begin{equation}
\frac{\dot{a}^2}{a^2} = \frac{8\pi G_N}{3} 
\left[\frac12 \dot{\phi}^2 + \frac12 \dot{\sigma}^2
+ \sum_{n=0}^{\infty} \frac{\sigma^{2n}}{2^n n!} 
V^{(2n)}(\phi)\right] \; .
\label{assembly3}
\end{equation}
Corrections to these effective classical equations due to modes
which have not yet crossed the horizon will be perturbatively 
small\cite{grav}.

Examination of the equations (\ref{assembly1}) -- (\ref{assembly3})
reveals an amazing result.  After a few $e$-folds of inflation, 
the full quantum evolution of spinodal
models of inflation for a real scalar inflaton may be modeled by 
a classical system of two homogeneous fields with potential
\begin{equation}
V(\phi,\sigma) = \sum_{n=0}^{\infty} \frac{1}{2^n n!} 
\sigma^{2n} V^{(2n)}(\phi) \; .
\label{2field}
\end{equation}

We emphasize that the two-field reassembled system arises from the Hartree
{\it dynamics} obtained by solving 
eqs.(\ref{phieqn}, \ref{fkeqn},\ref{frweqn}),
together with the initial conditions in eq.(\ref{fkin}). 

There are several important notes to make regarding this result.  First, the
initial value of $\sigma$ is not a free parameter; rather, it is determined
dynamically through the assembly process.  By completing the full quantum
evolution of $\langle \psi^2 \rangle$ until assembly is evident and then
extrapolating back to the initial time, we can show numerically that the
effective initial value of $\sigma$ is $\sigma(t_0) = {\cal F}(H_0)H_0/2\pi,$
where ${\cal F}(H_0)$ is a number of order $1$\cite{grav}.  Because the
effective initial value of $\sigma$ is fixed by the dynamics, there is a clear
separation into two regimes, depending upon the initial value of $\phi$.  The
first is the classical regime for which $\phi(t_0) \gg H_0/2\pi$; here the
dynamics will follow the usual classical dynamics for the original potential
$V(\phi)$. However, there exists a second regime with $\phi(t_0) < H_0/2\pi$
for which the quantum evolution modeled by the $\sigma$ field will have a
significant influence on the dynamics.

Second, while the dynamics described by the potential
(\ref{2field}) looks like that of a two field model of inflation, this is only
true in terms of the homogeneous dynamics; since the $\phi$ and $\sigma$ fields
are in fact just two aspects of the {\it same} scalar field, the computation of
the primordial power spectrum in these theories somewhat more subtle.

Now we turn to an example. Consider a spontaneously broken $\lambda
\Phi^4$ theory with a potential
\begin{equation}
V(\Phi)= \frac{3 m^4}{2 \lambda} - \frac{1}{2} m^2 \Phi^2 + \frac{\lambda}{4!}
\Phi^4.
\label{phi4pot}
\end{equation}
This potential vanishes at its minima at 
$\Phi = \pm \sqrt{6 m^2\slash \lambda}$.  
Of particular importance to us is the location of the
spinodal line separating the region of spinodal instability from
stability. This runs from $\Phi = -\sqrt{2 m^2\slash \lambda}$ to $\Phi =
+\sqrt{2 m^2\slash \lambda}$. In the large $N$ case treated in \cite{grav}, the
presence of Goldstone modes makes the spinodal line run along the {\it minima}
of the potential. 

If we evolve eqs.(\ref{phieqn}, \ref{fkeqn}, \ref{frweqn}) in time we arrive at
the results in fig.1.

The ``Hartreeized'' potential generated by the growth of quantum fluctuations
is 
\begin{equation}
V_{\rm hartree}(\phi, \sigma) = V(\phi)-\frac{\lambda}{4}(\phi_{\rm
spinodal}^2-\phi^2) \sigma^2 +\frac{\lambda}{8}\sigma^4,
\label{hartpot}
\end{equation}
where $\phi_{\rm spinodal}= \sqrt{2 m^2\slash \lambda}$.
\begin{figure}
\epsfig{file=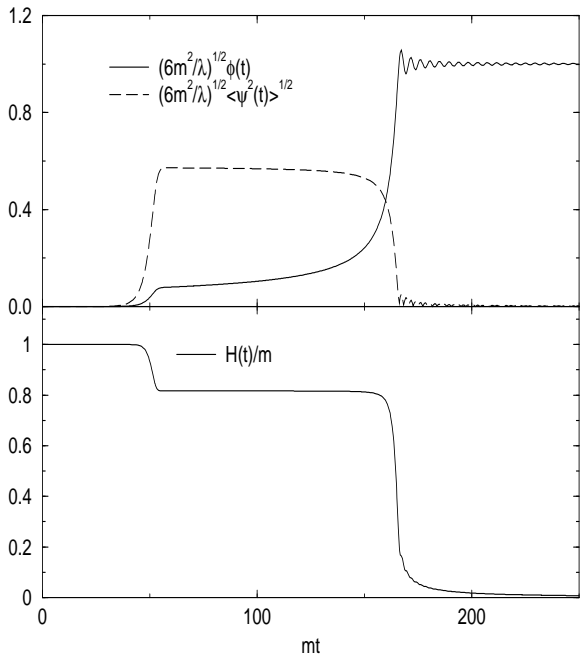,width=3in,height=3.5in}
\caption{Typical evolution of the zero mode $\phi$, the two-point 
fluctuation $\langle \psi^2 \rangle$, and the expansion rate $H \equiv
\dot{a}/a$ in the $\lambda \Phi^4$ theory.  Here, $\phi(t_0) = 0.2 H_0/m$,
$H_0/m=1$, and $\lambda=8\pi^2\times 10^{-14}$.}
\label{fig}
\end{figure}
We see that there are two inflationary stages. The first is driven by the
vacuum energy at $\phi\sim 0$. The fluctuations then grow and end that period
of inflation. However, once the fluctuations have grown large
enough, the second period of inflation ensues. Comparing the value of the
Hubble parameter $H$ in both phases, we see that the system now behaves as if
the order parameter was stuck near the {\it spinodal} value for the original
potential $V(\phi)$!
In fact, the dynamics is found to obey the following sum rule during this
second inflationary phase:
\begin{equation}
\langle \psi^2 \rangle = \left\{ \begin{array}{l}
\phi_{\rm spinodal}^2-\phi^2 \; \;  \mbox{if}\;  |\phi|\leq \phi_{\rm spinodal}\;, \\
0 \; \mbox{otherwise}
\end{array}
\right. 
\label{sumrule}
\end{equation}
In terms of the effective classical theory determined by the potential in
Eq.~(\ref{hartpot}), this sum rule corresponds to minimizing $V_{\rm
hartree}(\phi, \sigma)$ with respect to $\sigma$: $\partial_{\sigma}V_{\rm
hartree}(\sigma, \phi)=0$. Once $\phi$ reaches the spinodal, the minimizing
condition is that the fluctuations vanish and that $\phi$ evolves to the
minimum of the {\it tree-level} potential. In fact this is what the dynamics
shows. Note that the sum rule is reminiscent of the behavior of the
magnetization in a Heisenberg ferromagnet as a function of temperature. 

The number of e-folds of the second phase of inflation is determined by the
amount of time it takes $\phi$ to reach $\phi_{\rm spinodal}$; in the extreme
case that $\phi$ is fixed at $0$ this phase never ends. 

It is a worthwhile exercise to compare this behavior to the behavior found in
the large $N$ approximation. There, zero mode assembly also occured, but the
effective zero mode just evolved along the {\it classical} potential to the
tree-level minimum; there was no second inflationary phase. We would argue that
we are seeing the same behavior in both cases, the only difference being that
the existence of Goldstone modes in the large $N$ case makes the spinodal run
along the minima of the tree-level potential. Thus the reassembled zero mode
{\it does} in fact go to the spinodal line. 
However, in large $N$ that is just the
line of minima and since the potential is chosen to be zero at the minima,
there is no vacuum energy left to drive a second inflationary phase.

An alternative way to interpret what we are seeing here is that while field
fluctuations grow, they are producing particles whose effect is to modify the
background in which the field is evolving.  These particles are produced
copiously enough to survive the exponential redshift during the inflationary
phase. The ``potential'' that the zero mode will follow is now one that must be
computed in the presence of the bath of produced particles. Once the zero mode
crosses the spinodal line, however, particle production ends and their effect
is obliterated by redshifting, thus allowing the zero mode to find the minimum
of the tree-level potential.

Finally, we address the issue of metric perturbations in spinodally unstable
theories.  There are two aspects of the evolution which will result in
departures from the conventional wisdom for this class of models.  First is the
two stage nature of the evolution which may result in a deviation from scale
invariance and a kink in the power specturm\cite{liddle}.  Second is the
effective two field dynamics in which the zero mode slowly evolves to the
spinodal line and triggers the end of inflation.  This is reminiscent of hybrid
inflation models and may have similar consequences, such as the production of a
blue tilt to the power spectrum\cite{hybrid}. Any of these possibilities will
have significant consequences for the reconstruction program\cite{recon}.  A
proper study using the gauge invariant formalism\cite{mfb} is in
progress\cite{us}.

In this work we have made heavy use of the Hartree approximation. It is
reasonable to ask whether this approximation captures the essential physics of
the situation. First we should note that, unlike the large $N$ case, the
Hartree approximation is a {\it truncation} of the theory that is uncontrolled
in the sense that we do not have a way to systematically go beyond it. There is
some hope that the 2PI formalism of Cornwall, Jackiw and Tomboulis\cite{cjt},
could be used to at least try to estimate the diagrams omitted in the Hartree
truncation\cite{us}.

However, we should be heartened by the fact that the Hartree approximation {\it
does} recognize the importance of the spinodal line. As shown by Weinberg and
Wu\cite{weinwu}, the existence of the spinodal line is correlated both with the
non-convexity as well as with the imaginary part of the one-loop effective
potential. This imaginary part corresponds to the decay rate of a state
prepared so that the zero mode is localized near the top of the potential. Thus
spinodal models are {\it inherently} dynamical and cannot be treated within the
confines of the effective potential approximation. What the Hartree
approximation provides us with is a way to deal with spinodal dynamics beyond
perturbation theory.

The result is a dynamical evolution in which non-perturbative
{\it quantum} fluctuations play a primary role in the evolution of the
inflaton field, the gravitational background, and the production
of primordial metric perturbations.

\acknowledgements 
The authors would like to thank Dan Boyanovsky and Ed Copeland for
important contributions to this work.
D.C. was supported by the North Atlantic Treaty Organization under
NSF grant DGE-98-04564.  R.H. was supported by DOE grant
DE-FG02-91-ER40682.

\end{document}